\PassOptionsToPackage{pdfpagelabels=false}{hyperref}

\documentclass{IOS-Book-Article}

\usepackage{mathptmx}
\usepackage{soul}\setuldepth{article}
\usepackage{graphicx}
\usepackage{hyperref}
\def\hb{\hbox to 11.5 cm{}}

\begin{document}

\pagestyle{headings}
\def\thepage{}
\begin{frontmatter}              

\title{Intention and Context Elicitation with Large Language Models in the Legal Aid Intake Process}

\markboth{}{November 2023\hb}

\author[A]{\fnms{Nick} \snm{Goodson}
\thanks{Contact authors at ngoodson@alumni.stanford.edu and rongfeil@stanford.edu.}}
and
\author[A]{\fnms{Rongfei} \snm{Lu}}
\address[A]{Stanford University}

\begin{abstract}

Large Language Models (LLMs) and chatbots show significant promise in streamlining the legal intake process. This advancement can greatly reduce the workload and costs for legal aid organizations, improving availability while making legal assistance more accessible to a broader audience. However, a key challenge with current LLMs is their tendency to overconfidently deliver an immediate 'best guess' to a client's question based on the output distribution learned over the training data. This approach often overlooks the client's actual intentions or the specifics of their legal situation. As a result, clients may not realize the importance of providing essential additional context or expressing their underlying intentions, which are crucial for their legal cases. Traditionally, logic based decision trees have been used to automate intake for specific access to justice issues, such as immigration and eviction. But those solutions lack scalability. We demonstrate a proof-of-concept using LLMs to elicit and infer clients' underlying intentions and specific legal circumstances through free-form, language-based interactions. We also propose future research directions to use supervised fine-tuning or offline reinforcement learning to automatically incorporate intention and context elicitation in chatbots without explicit prompting.

\end{abstract}

\end{frontmatter}
\markboth{November 2023\hb}{November 2023\hb}

\section{Introduction}

Legal aid centers and courts traditionally depend on human experts and attorneys to gather information from clients, assess each case, and guide clients to appropriate legal resources and services. These experts also advise clients on developing legal strategies, navigating court processes, filling out forms, preparing for hearings, and handling other legal tasks \cite{hagan_towards_2023}. In the United States, where access to justice is a significant concern, machine learning systems, particularly Large Language Models (LLMs), hold great promise for enhancing service quality and outreach for clients in need.

State of the art LLMs such as OpenAI's GPT-4 \cite{openai_gpt-4_2023} have been trained on a vast corpus of human knowledge. Given a sufficiently constrained legal question, such models can often respond with accuracy comparable to that of legal professionals, as demonstrated by \cite{katz_gpt-4_2023}. However, the legal intake process remains challenging due to two main factors:

\begin{enumerate}

\item Clients, typically without legal expertise, may ask questions that do not fully reveal their underlying intentions or needs. Every client aims to achieve a specific objective, but their limited knowledge can lead them to pose questions that do not fully embody this objective. This issue is especially pronounced in the access to justice context, where clients may have had minimal or no prior engagement with the legal system. As a result, they are likely to inadvertently ask the "wrong" questions, failing to elicit the most useful answers for their situation.

\item The most effective responses to client legal questions are contingent on the specifics of each client's circumstances. However, the client may not know which contextual details to include in their query. In the baseline approach, an LLM samples a 'best guess' answer from the underlying distribution of outputs using only the information presented. Typically this does not result in the model probing for additional, and often crucial, context. Therefore, it may fall short in addressing the nuanced and specific needs of individual legal cases.

\end{enumerate}

In this paper, we propose a framework designed to adapt LLMs for more effective use in legal aid contexts. Our approach focuses on enabling these models to actively seek a comprehensive understanding of a client’s situation and to draw out their underlying intentions. By integrating this contextual information as an additional input, the LLM can generate more accurate and relevant responses. This process is achieved without the need for further fine-tuning of the model, relying instead on its ability to interpret and apply the gathered context to enhance its output.


\section{Related Works}

\subsection{Artificial Intelligence to increase Access to Justice}

Individuals around the world face legal problems that impact fundamental aspects of their lives, such as housing, finances, citizenship, employment, and family. Issues like eviction, workplace harassment, domestic violence, child custody, political asylum and accessing government benefits are common \cite{hagan_towards_2023}. Often people struggle to resolve their issues through the legal system, either due to a lack of awareness that legal assistance is necessary or difficulties in accessing suitable legal resources. This challenge is compounded by the limited capacity of legal aid organizations and court centers. 

Several researchers have explored solutions to this problem. Thomson introduced the Justice Pathway Expert System (JPES), a conceptual description of an expert system designed to improve access to justice for laypersons. JPES would use expert knowledge from various legal fields to diagnose problems, deliver tailored information, provide self-help support, and guide users through Online Dispute Resolution (ODR) processes \cite{thompson_creating_2015}. Zeleznikow developed the GetAid system, aimed at assisting lawyers in assessing individual's eligibility for legal aid \cite{zeleznikow_using_2002}. The Dutch legal aid board created Rechtwijzer 1.0 and 2.0 platforms, which gather users' situational data and direct them to relevant resources, particularly in consumer disputes and divorce \cite{bickel_online_2015}.

This paper contributes to this line of research by presenting a novel AI research direction. Our aim is to develop methods that effectively gather contextual information and client goals. This approach will facilitate efficient triage and consultation on legal problems without the need for creating specific rules and expert systems for each issue.

\subsection{Rule-based systems}

A notable approach in legal research involves translating laws and statutes into rule-based systems. This method simplifies the legal texts, making them more accessible for automated analysis and application. For instance, Allen and Engholm converted legal statutes into propositional logic, enhancing their clarity and interpretability \cite{allen_normalized_1978}. Sergot et al adopted a similar approach, implementing the British Nationality Act into the Prolog programming language, encompassing around 150 rules \cite{sergot_british_1986}. Walker explored the idea of modeling legislation as decision trees, applying this concept to Board of Veterans' appeal cases \cite{walker_semantic_2017}. We also see similar applications in landlord-tenant disputes \cite{westermann_using_2019}, among others.

\subsection{Case-based systems}

Another common approach involves using historical case data to enhance legal reasoning and prediction. An early example of this approach is the HYPO expert system developed by Ashley. HYPO utilized commonly observed fact patterns that influence the strength of a plaintiff’s argument in trade secret law \cite{ashley_reasoning_1991}. Subsequent research built on this foundation, such as the Value Judgement-based Argumentative Prediction (VJAP). This system linked factors to values, allowing for reasoning based on policy balances and generating comprehensive argument structures \cite{grabmair_predicting_2017}. Further advancements include the work of \cite{al-abdulkarim_methodology_2016}, which employed abstract dialectic frameworks to represent rules derived from cases. This methodology enables the prediction of outcomes in new cases and the generation of explanations. Additionally, \cite{westermann_justicebot_2023} proposed a hybrid system that combines rule-based and case-based reasoning, integrating the strengths of both approaches.

Machine learning has been increasingly utilized in legal research to develop models for legal reasoning and to predict case outcomes or trends. Ashley pioneered the Issue-based Prediction (IBP) model, which leverages domain-specific data in conjunction with case information to predict outcomes \cite{ashley_automatically_2009}. Yin et al used supervised learning to build an extended multi-layer perceptron neutral network. This model classifies individuals as employers or independent contractors, achieving a high accuracy \cite{yin_determining_2020}. Branting et al introduced two novel approaches for explainable legal outcome prediction. One uses an attention network to highlight salient case text and the other uses semi-supervised case annotation for generating legal explanations \cite{branting_scalable_2020}. \cite{westermann_justicebot_2023} annotated real-life decisions in landlord-tenant disputes and used them to predict outcomes and discover trends. Stranieri et al explored automated legal reasoning in the discretionary domain. They built a system for predicting property distribution in Australian divorce cases by integrating neural networks with rule-based reasoning \cite{stranieri_hybrid_1999}. Meanwhile, Dahan et al compared various machine learning techniques such as regression, decision tree, random forests and neural networks to predict notice periods following employment termination. However, they faced challenges in achieving high accuracy \cite{dahan_predicting_2020}.

\subsection{Large Language Models}

Recent advancements in Large Language Models (LLMs) such as GPT-3 and BERT have demonstrated human-like proficiency in language understanding and generation. Studies have underscored the versatility of these models, showing their effectiveness in various applications like text completion, language translation, creative content generation, and emergent general intelligence without explicit prompting \cite{bubeck_sparks_2023}. Legal experts believe that tools like ChatGPT and other AI chatbots hold significant potential to increase access to justice \cite{greggwirth_forum_2023}\cite{holt_legal_nodate}.

However, it is acknowledged that ChatGPT and similar tools do not always provide accurate or reliable information. Despite this, they offer a user-friendly way to engage with laypeople \cite{westermann_using_2019}. Hagan's user interviews and empirical studies suggest a growing preference for using AI tools like Bard or ChatGPT in addressing legal issues. She hypothesizes that people will increasingly rely on LLMs for handling legal problems \cite{hagan_towards_2023}. Yet, there's a risk of over-dependence on LLMs for legal explanations, despite their limitations such as hallucinating legal cases, referencing inapplicable organizations, providing incorrect jurisdictional laws and procedures, which may lead to users cherry-picking legal details \cite{hagan_towards_2023}.

Our paper addresses these challenges by proposing a methodology for enhancing the accuracy and relevance of LLMs in legal settings. We focus on improving context and intention understanding from clients, aiming to reduce biases and enhance the trustworthiness and applicability of these powerful AI tools in legal consultations.

\subsection{LLM-based Approach in Intention and Context Elicitation}

Intention and context elicitation is vital not only in legal settings but also in \textit{general goal-directed conversations} \cite{hong_zero-shot_2023}. For instance, an LLM can be prompted to act as a travel agent and even produce realistic two-sided conversations; however, it will not intentionally try to maximize the probability of planning a desirable itinerary for a real human \cite{hong_zero-shot_2023}.  This often results in conversations lacking in clarifying questions, goal-directed flow, and personalization. Outputs tend to be verbose and non-specific. Hong et al suggested using LLMs to generate possible goal-directed dialogue examples, then applying offline reinforcement learning to process these examples and learn to perform more optimal interactions \cite{hong_zero-shot_2023}. Similarly, Li and Tamkin  leveraged LLMs to elicit human preferences in three task domains - email validation, content recommendation, and moral reasoning. Their findings indicate that responses elicited by LLMs are typically more informative than user-written prompts or labels. Users also report that interactive task elicitation through LLMs requires less effort than prompting or example labeling and surfaces novel considerations not initially anticipated by users \cite{li_eliciting_2023}. Our system adopts similar strategies to elicit critical information from legal clients, enhancing the efficiency and effectiveness of legal consultations through AI.

\section{LLMs for Legal Intake}

This section presents a proof-of-concept design for an automated legal intake system built upon LLMs. The most straightforward approach—directly inputting a client's inquiry into an LLM—proves to have limited utility. This is primarily because it depends on the client providing all relevant information upfront, a method further discussed in section \ref{section:one-shot}.

A more effective design involves eliciting additional contextual information after receiving the initial client inquiry. This leads to the adoption of a multi-step conversational model, mirroring the process of intake with a human attorney. The primary objective is to engage in a conversation to gather relevant additional information before addressing the client’s legal question directly. This approach offers two main advantages. Firstly, it creates a more comprehensive case profile that can be passed on to human attorneys. Secondly, it enhances the model’s capability to provide accurate initial guidance to the client, even before they consult an attorney. This aspect is increasingly important, considering the growing trend of individuals using LLMs as a search tool for personal legal issues \cite{hagan_towards_2023}.

While this approach shows initial promise, it is important to note that the results at this stage are qualitative and do not represent the findings of extensive experimental research.

\subsection{One-Shot Responses}
\label{section:one-shot}
\begin{figure}
    \centering
    \includegraphics[width=0.5\textwidth]{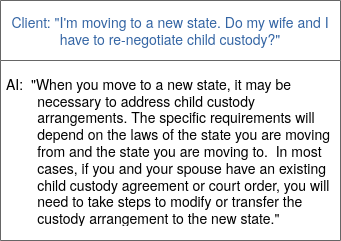}
    \caption{One-shot LLM model output example demonstrating the generic responses returned for vague client questions.}
    \label{fig:One-shot-answers}
\end{figure}

The simplest method for legal intake with Large Language Models (LLMs) is one-shot question answering. This technique involves priming the model with a general background context before presenting the client's question directly. When the client's question is well-crafted, this approach can provide useful results, as the legal capabilities of LLMs have been well established \cite{bubeck_sparks_2023}.

However, the effectiveness of this method declines sharply with questions that lack specificity or omit crucial legal details. In such cases, the inherent overconfidence of LLMs often results in responses that are either vague or too generic to be of practical use to the client. An illustrative example from the family law domain, shown in \textbf{Figure \ref{fig:One-shot-answers}}, demonstrates this issue. Here, the model's response is overly broad concerning state laws and fails to ascertain and address whether the client wishes to renegotiate child custody.

\subsection{Intention and Context Elicitation}

To improve the automation of legal intake processes using Large Language Models (LLMs), it's crucial for models to identify and seek out unknown information. This information can be broadly categorized into two types: Intentions and Context. Once this information is elicited from the client, it can be combined with their original question to better prime the model for generating a response that is most beneficial for the client.

\subsubsection{Intentions}

\begin{figure}
    \centering
    \includegraphics[width=0.95\textwidth]{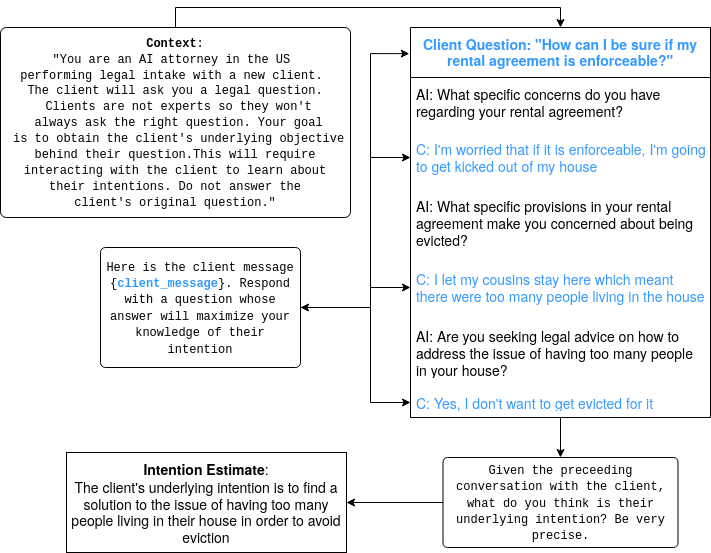}
    \caption{The intention-elicitation model structure including prompts and an example tenancy law question. The generated intention estimate is the primary output.}
    \label{fig:intention-convo}
\end{figure}

Directly interpreting a client’s underlying intention from their initial question can be challenging. For instance, in immigration law, a client may inquire about a specific visa, but their actual intention could be to legally work near their partner in the U.S. There might be a more suitable visa option for the client than the one they are considering, revealing a gap due to their limited knowledge.

Our model for eliciting intentions is inspired by the human legal-intake process. It involves instructing the LLM to engage in a conversation with the client, focusing on two primary inputs. First, the model is primed with the goal of uncovering the client's underlying intention, rather than immediately answering the initial question. Second, each client input, including the initial question, is contextualized in a way that prompts the model to ask questions that maximize the gathering of relevant information. \textbf{Figure \ref{fig:intention-convo}} illustrates an example conversation, detailing the prompts used to facilitate it.

This approach enables the model to reveal insights fundamental to the legal case. At the conversation's conclusion, the model synthesizes its best estimate of the client's underlying intention. This estimate, combined with the elicited context described in section \ref{section:context}, primes the model to produce a final, informed response to the client's query. An example of such an intention estimate is presented in \textbf{Figure \ref{fig:intention-convo}}.

\subsubsection{Context}
\label{section:context}
\begin{figure}
    \centering
    \includegraphics[width=0.95\textwidth]{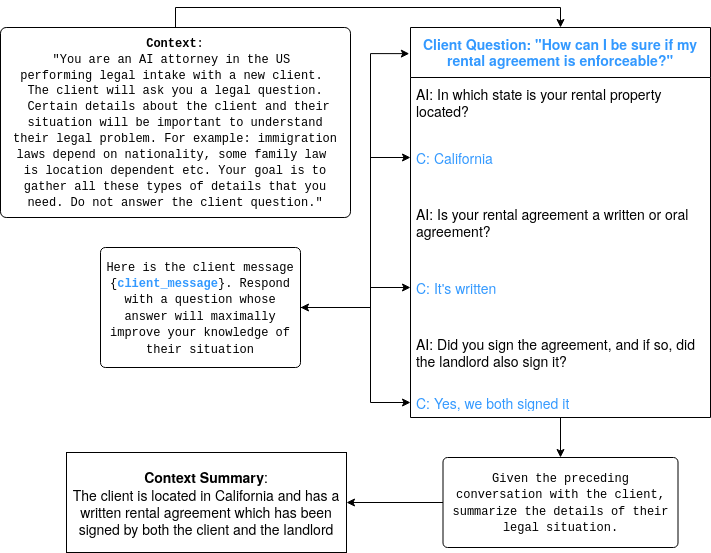}
    \caption{The context-elicitation model structure including prompts and the same tenancy law example as in Figure \ref{fig:intention-convo}. The generated context summary is the primary output}
    \label{fig:context-convo}
\end{figure}

Contextual information in a legal setting pertains to factual details specific to a client’s situation. For instance, in immigration law, certain regulations depend on the client's nationality. Similarly, in real estate and tenancy law, the applicable laws vary by location. The process of eliciting this context mirrors the previously described intention-elicitation method, involving a conversational approach with the client.

In this process, the LLM is primed with a specific objective: to gather all legally relevant information about the client’s situation. Each input from the client is framed within a context that prompts the model to ask questions aimed at maximizing the acquisition of pertinent details about the client's circumstances. An example of this context-elicitation in action, using the same family law scenario, is illustrated in \textbf{Figure \ref{fig:context-convo}}.

Just like with elicited intentions, the model is instructed to summarize the contextual information it has gathered. This summary is then utilized to prime the model’s final response to the client’s query.

\subsection{Combined Output}

The information obtained through intention and context elicitation can be highly beneficial for human attorneys in understanding a client's needs. Additionally, this information can be used to generate immediate responses for clients. These responses are tailored to address the client's underlying intentions, taking into account the contextual details of their situation as well as their original question.

\begin{figure}
    \centering
    \includegraphics[width=\textwidth]{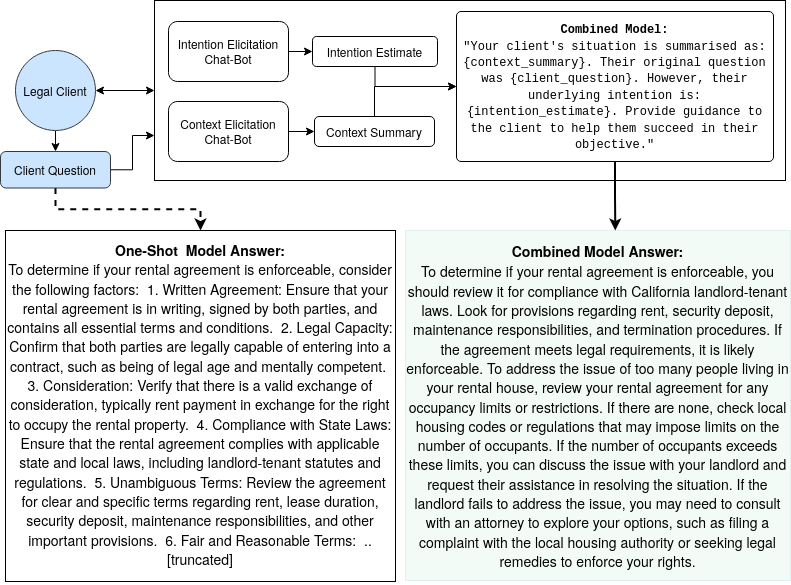}
    \caption{The outputs of the intention-elicitation and context-elicitation sub-models can be combined to provide an immediate response to the client. This response provides qualitatively provides higher utility than the one-shot response shown. }
    \label{fig:combined-output}
\end{figure}

\textbf{Figure \ref{fig:combined-output}} showcases how the integration of these three elements—intention, context, and the original question—results in a response that is useful for the client. For comparison, the figure also includes a one-shot response to the same query. In contrast to the nuanced, tailored response from our method, the one-shot method yields a generic checklist of legal specifics. These are often hard to comprehend and may not be relevant to the client’s specific municipal laws. The combined answer gives a list of specific suggestions tailored to the client's legal circumstances. 

\subsection{Selecting Sub-Components}

\begin{figure}
    \centering
    \includegraphics[width=0.5\textwidth]{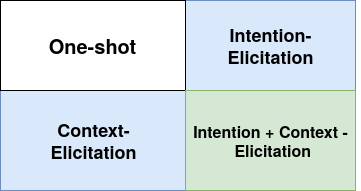}
    \caption{Combinations of LLM legal-intake model components}
    \label{fig:component-combos}
\end{figure}

\textbf{Figure \ref{fig:component-combos}} illustrates the various configuration options for the LLM-based legal intake system. Based on our initial qualitative assessments, we estimate that the best results can be achieved using a combined approach of both intention and context elicitation.

However, it's important to consider scenarios where using just one of these components might be more appropriate. For instance, if a client's question clearly conveys their underlying intention, employing the intention-elicitation model might unnecessarily prolong the intake process. Similarly, if a client provides a detailed account of their situation, the context-elicitation model might not add significant value.

One might consider pre-filtering questions with an LLM to determine the necessity of each component. The model could be prompted to assess if the client’s question contains enough information for a complete answer. However, creating a prompt that reliably controls this behavior is challenging. There is no robust metric to define what constitutes a complete answer. Moreover, the overconfidence of LLMs often leads to a high rate of false positives, where the model assumes it has sufficient information even when significant ambiguity exists.

Given these considerations, it seems most practical to apply the full method, encompassing both intention and context elicitation, to every client question.

\section{Discussions and Conclusion}

In this paper, we introduced a proof-of-concept utilizing Large Language Models (LLMs) to gain a detailed understanding of clients’ situations and elicit their underlying intentions. Further, we showed how this information can be used as input to an LLM to generate qualitatively better final responses without the need for additional fine-tuning. We propose that LLMs equipped with the capability to actively solicit client intentions and relevant context can significantly enhance access to justice by improving the efficiency of legal intake and triage processes in legal aid and court centers.

\subsection{Limitations}

This paper represents an initial exploration into the use of intention and context elicitation with LLMs in legal intake, aimed at advancing access to justice. The next steps involve conducting experiments and ablation studies with both machine and human evaluators. These studies will compare the effectiveness of LLMs in soliciting intention and context against the baseline one-shot answers. 

\subsection{Ethical Considerations}  

Echoing concerns raised by Hagan \cite{hagan_towards_2023}, there is a tendency for clients to over-rely on information provided by LLMs, despite its potential inaccuracies. In a production environment, it is essential for responses generated by LLMs to be reviewed by attorneys. This step not only addresses issues of model hallucination but also adds a necessary human element for clients. It may also help mitigate concerns about the unauthorized practice of law. Additionally, clients should be informed about and consent to the use of AI, with a strong emphasis on the protection of their privacy. Following the ethical use of AI guidelines by the California Bar Association \cite{california_bar} and American Bar Association, real client data and personal information should not be used to train AI models. Such AI tools should also comply with California Consumer Privacy Act (CCPA) and other applicable data privacy protection laws to ensure client privacy.

\subsection{Future Directions}

In the future, we aim to use the datasets generated from intention and context elicitation conversations with LLMs to train an interactive conversational agent. This agent will optimize goal-directed objectives over multiple interactions, using methods such as supervised fine-tuning or offline reinforcement learning. We also plan to conduct experiments to showcase potential performance enhancements over standard zero-shot answers. Furthermore, we intend to release LLM-generated conversation datasets, verified by attorneys, that encompass various personas and issues pertinent to the most critical legal domains in the access to justice space.

\bibliographystyle{vancouver}
\bibliography{elicitation_a2j}

\begin{thebibliography}{10}

\bibitem{hagan_towards_2023}
Hagan M.
\newblock Towards {Human}-{Centered} {Standards} for {Legal} {Help} {AI} [{SSRN} {Scholarly} {Paper}].
\newblock Philosophical Transactions of the Royal Society A: Mathematical, Physical and Engineering Sciences. 2023 Sep.
\newblock Available from: \url{https://papers.ssrn.com/abstract=4582745}.

\bibitem{openai_gpt-4_2023}
OpenAI.
\newblock {GPT}-4 {Technical} {Report}.
\newblock OpenAI; 2023.
\newblock ArXiv:2303.08774 [cs].
\newblock Available from: \url{http://arxiv.org/abs/2303.08774}.

\bibitem{katz_gpt-4_2023}
Katz DM, Bommarito MJ, Gao S, Arredondo P.
\newblock {GPT}-4 {Passes} the {Bar} {Exam}.
\newblock Rochester, NY: Stanford CodeX Center for Legal Informatics; 2023.
\newblock Available from: \url{https://papers.ssrn.com/abstract=4389233}.

\bibitem{thompson_creating_2015}
Thompson D.
\newblock Creating {New} {Pathways} to {Justice} {Using} {Simple} {Artificial} {Intelligence} and {Online} {Dispute} {Resolution} [{SSRN} {Scholarly} {Paper}].
\newblock International Journal of Online Dispute Resolution. 2015 Jun.
\newblock Available from: \url{https://papers.ssrn.com/abstract=2696499}.

\bibitem{zeleznikow_using_2002}
Zeleznikow J.
\newblock Using {Web}-based {Legal} {Decision} {Support} {Systems} to {Improve} {Access} to {Justice}.
\newblock Information \& Communications Technology Law. 2002 May.
\newblock Publisher: Taylor \& Francis Group.
\newblock Available from: \url{https://www.tandfonline.com/doi/abs/10.1080/13600830220133530}.

\bibitem{bickel_online_2015}
Bickel EA, Dijk MAJv, Giebels E.
\newblock Online legal advice and conflict support: a {Dutch} experience.
\newblock Department Psychology of Conflict, Risk \& Safety, University of Twente; 2015.

\bibitem{allen_normalized_1978}
Allen L, Engholm C.
\newblock Normalized {Legal} {Drafting} and the {Query} {Method}.
\newblock Articles. 1978 Jan.
\newblock Available from: \url{https://repository.law.umich.edu/articles/29}.

\bibitem{sergot_british_1986}
Sergot MJ, Sadri F, Kowalski RA, Kriwaczek F, Hammond P, Cory HT.
\newblock The {British} {Nationality} {Act} as a logic program.
\newblock Communications of the ACM. 1986 May;29(5):370-86.
\newblock Available from: \url{https://dl.acm.org/doi/10.1145/5689.5920}.

\bibitem{walker_semantic_2017}
Walker V, Okpara N, Hemendinger A, Ahmed T.
\newblock Semantic {Types} for {Decomposing} {Evidence} {Assessment} in {Decisions} on {Veterans}’ {Disability} {Claims} for {PTSD}.
\newblock Hofstra Law Faculty Scholarship. 2017 Jun.
\newblock Available from: \url{https://scholarlycommons.law.hofstra.edu/faculty_scholarship/1134}.

\bibitem{westermann_using_2019}
Westermann H, Walker VR, Ashley KD, Benyekhlef K.
\newblock Using {Factors} to {Predict} and {Analyze} {Landlord}-{Tenant} {Decisions} to {Increase} {Access} to {Justice}.
\newblock In: Proceedings of the {Seventeenth} {International} {Conference} on {Artificial} {Intelligence} and {Law}. {ICAIL} '19. New York, NY, USA: Association for Computing Machinery; 2019. p. 133-42.
\newblock Available from: \url{https://dl.acm.org/doi/10.1145/3322640.3326732}.

\bibitem{ashley_reasoning_1991}
Ashley KD.
\newblock Reasoning with cases and hypotheticals in {HYPO}.
\newblock International Journal of Man-Machine Studies. 1991 Jun;34(6):753-96.
\newblock Available from: \url{https://www.sciencedirect.com/science/article/pii/002073739190011U}.

\bibitem{grabmair_predicting_2017}
Grabmair M.
\newblock Predicting trade secret case outcomes using argument schemes and learned quantitative value effect tradeoffs.
\newblock In: Proceedings of the 16th edition of the {International} {Conference} on {Articial} {Intelligence} and {Law}. {ICAIL} '17. New York, NY, USA: Association for Computing Machinery; 2017. p. 89-98.
\newblock Available from: \url{https://dl.acm.org/doi/10.1145/3086512.3086521}.

\bibitem{al-abdulkarim_methodology_2016}
Al-Abdulkarim L, Atkinson K, Bench-Capon T.
\newblock A methodology for designing systems to reason with legal cases using {Abstract} {Dialectical} {Frameworks}.
\newblock Artificial Intelligence and Law. 2016 Mar;24(1):1-49.
\newblock Available from: \url{https://doi.org/10.1007/s10506-016-9178-1}.

\bibitem{westermann_justicebot_2023}
Westermann H, Benyekhlef K.
\newblock {JusticeBot}: {A} {Methodology} for {Building} {Augmented} {Intelligence} {Tools} for {Laypeople} to {Increase} {Access} to {Justice}.
\newblock In: Proceedings of the {Nineteenth} {International} {Conference} on {Artificial} {Intelligence} and {Law}. {ICAIL} '23. New York, NY, USA: Association for Computing Machinery; 2023. p. 351-60.
\newblock Available from: \url{https://doi.org/10.1145/3594536.3595166}.

\bibitem{ashley_automatically_2009}
Ashley KD, Brüninghaus S.
\newblock Automatically classifying case texts and predicting outcomes.
\newblock Artificial Intelligence and Law. 2009 Jun;17(2):125-65.
\newblock Available from: \url{https://doi.org/10.1007/s10506-009-9077-9}.

\bibitem{yin_determining_2020}
Yin Y, Zulkernine F, Dahan S.
\newblock Determining {Worker} {Type} from {Legal} {Text} {Data} using {Machine} {Learning}.
\newblock In: 2020 {IEEE} {Intl} {Conf} on {Dependable}, {Autonomic} and {Secure} {Computing}, {Intl} {Conf} on {Pervasive} {Intelligence} and {Computing}, {Intl} {Conf} on {Cloud} and {Big} {Data} {Computing}, {Intl} {Conf} on {Cyber} {Science} and {Technology} {Congress} ({DASC}/{PiCom}/{CBDCom}/{CyberSciTech}); 2020. p. 444-50.
\newblock Available from: \url{https://ieeexplore.ieee.org/document/9251195}.

\bibitem{branting_scalable_2020}
Branting LK, Pfeifer C, Brown B, Ferro L, Aberdeen J, Weiss B, et~al.
\newblock Scalable and {Explainable} {Legal} {Prediction}.
\newblock Artificial Intelligence and Law. 2020;29(2):213-38.
\newblock Publisher: Springer Verlag.

\bibitem{stranieri_hybrid_1999}
Stranieri A, Zeleznikow J, Gawler M, Lewis B.
\newblock A hybrid rule – neural approach for the automation of legal reasoning in the discretionary domain of family law in {Australia}.
\newblock Artificial Intelligence and Law. 1999 Sep;7(2):153-83.
\newblock Available from: \url{https://doi.org/10.1023/A:1008325826599}.

\bibitem{dahan_predicting_2020}
Dahan S, Touboul J, Lam J, Sfedj D.
\newblock Predicting {Employment} {Notice} {Period} with {Machine} {Learning}: {Promises} and {Limitations} [{SSRN} {Scholarly} {Paper}].
\newblock McGill Law Journal / Revue de droit de McGill. 2020 May.
\newblock Available from: \url{https://papers.ssrn.com/abstract=3595769}.

\bibitem{bubeck_sparks_2023}
Bubeck S, Chandrasekaran V, Eldan R, Gehrke J, Horvitz E, Kamar E, et~al.
\newblock Sparks of {Artificial} {General} {Intelligence}: {Early} experiments with {GPT}-4.
\newblock Microsoft Research; 2023.
\newblock ArXiv:2303.12712 [cs].
\newblock Available from: \url{http://arxiv.org/abs/2303.12712}.

\bibitem{greggwirth_forum_2023}
greggwirth.
\newblock Forum: {There}’s potential for {AI} chatbots to increase access to justice.
\newblock Thomson Reuters Institute; 2023.
\newblock Available from: \url{https://www.thomsonreuters.com/en-us/posts/legal/forum-spring-2023-ai-chatbots/}.

\bibitem{holt_legal_nodate}
Holt A.
\newblock Legal {AI}-d to {Your} {Service}: {Making} {Access} to {Justice} a {Reality}.
\newblock Vanderbilt University; 2023.
\newblock Available from: \url{https://www.vanderbilt.edu/jetlaw/2023/02/04/legal-ai-d-to-your-service-making-access-to-justice-a-reality/}.

\bibitem{hong_zero-shot_2023}
Hong J, Levine S, Dragan A.
\newblock Zero-{Shot} {Goal}-{Directed} {Dialogue} via {RL} on {Imagined} {Conversations}.
\newblock UC Berkeley; 2023.
\newblock ArXiv:2311.05584 [cs].
\newblock Available from: \url{http://arxiv.org/abs/2311.05584}.

\bibitem{li_eliciting_2023}
Li BZ, Tamkin A, Goodman N, Andreas J.
\newblock Eliciting {Human} {Preferences} with {Language} {Models}.
\newblock MIT, Stanford; 2023.
\newblock ArXiv:2310.11589 [cs].
\newblock Available from: \url{http://arxiv.org/abs/2310.11589}.

\bibitem{california_bar}
Committee~on Professional~Responsibility SBoC.
\newblock Practical Guidance for the Use of Generative Artificial Intelligence in the Practice of Law.
\newblock the State Bar of California; 2023.
\newblock Available from: \url{https://www.calbar.ca.gov/Attorneys/Conduct-Discipline/Ethics/Ethics-Technology-Resources}.

\end{thebibliography}
\end{document}